\documentclass[11pt,prd,aps,showpacs,eqsecnum,floatfix,nofootinbib,preprint,tightenlines]{revtex4}
\usepackage{graphicx}				
\usepackage{latexsym}
\usepackage{epsf}
\usepackage{graphicx}
\usepackage{slashed}
\usepackage{amsmath}
\usepackage{amssymb}

\usepackage{color}
\usepackage{colordvi}
\usepackage{multirow}
\newcommand{\be}{\begin{equation}}
\newcommand{\ee}{\end{equation}}
\newcommand{\bea}{\begin{eqnarray}}
\newcommand{\eea}{\end{eqnarray}}
\newcommand{\ba}{\begin{array}}
\newcommand{\ea}{\end{array}}
\newcommand{\nn}{\nonumber}

\newcommand{\pref}[1]{(\ref{#1})}
\newcommand{\np}{N\pi}

\newcommand{\ENq}{E_{N,\vec{q}}}

\newcommand{\Epiq}{E_{\pi,{\vec{q}}}}
\newcommand{\Epip}{\Epik}

\newcommand{\Epik}{E_{\raisebox{0.05ex}[0pt]{$\scriptstyle \pi,\vec{k}$}}}

\newcommand{\GA}{G_{\rm A}}
\newcommand{\GP}{G_{\rm P}}
\newcommand{\GPt}{\tilde{G}_{\rm P}}

\newcommand{\epsA}{\epsilon_{\rm A}}
\newcommand{\epsPt}{\epsilon_{\rm \tilde{P}}}
\newcommand{\epsP}{\epsilon_{\rm P}}
\newcommand{\DG}{\Delta G}

\begin{document}
\renewcommand{\thefootnote}{$*$}

\preprint{HU-EP-19/15}

\title{$N\pi$-state contamination in lattice calculations of the nucleon pseudoscalar form factor}

\author{Oliver B\"ar$^{a}$} 
\affiliation{$^a$Institut f\"ur Physik,
\\Humboldt Universit\"at zu Berlin,
\\12489 Berlin, Germany\\}

\begin{abstract}
The nucleon-pion-state contribution in the QCD 3-point function of the pseudoscalar density is calculated to leading order in chiral perturbation theory. It predicts a nucleon-pion-state contamination in lattice estimates for the pseudoscalar form factor $\GP(Q^2)$ determined with the plateau method. Depending on the momentum transfer $Q^2$ the contamination varies between -20\% and +50\% for a source-sink separation  of 2 fm. The nucleon-pion-state contamination also causes  violations in the generalized Goldberger-Treiman relation among the pseudoscalar and the axial nucleon form factors, the dominant source being the nucleon-pion-state contamination in the induced pseudoscalar form factor $\GPt(Q^2)$.
Comparing the chiral perturbation theory predictions with lattice results of the PACS collaboration we find reasonable agreement even for source-sink separations as small as 1.3 fm. 
\end{abstract}

\maketitle

\renewcommand{\thefootnote}{\arabic{footnote}} \setcounter{footnote}{0}

\newpage

%
\section{Introduction}
%
In a recent paper \cite{Bar:2018xyi} the nucleon-pion ($N\pi$) contribution in the nucleon axial vector 3-point (pt) function was computed in leading order (LO) chiral perturbation theory (ChPT). The results allow to estimate the $N\pi$-state contamination in lattice QCD estimates for the two associated axial form factors, the axial form factor $\GA(Q^2)$ and the induced pseudoscalar form factor $\GPt(Q^2)$. In particular the latter is afflicted with a sizeable $N\pi$-state contamination leading  to an underestimation of the true form factor for small momentum transfers $Q^2$. As a result the $Q^2$ dependence of the form factor is distorted and differs significantly from the one expected by the pion-pole dominance (PPD) model.  

Due to the partially conserved axial vector current (PCAC) relation the 3-pt function of the axial vector current is related to the 3-pt function of the pseudoscalar density. This relation implies a relation between the two axial form factors and the pseudoscalar form factor $\GP(Q^2)$, often called the generalized Goldberger-Treiman (gGT) relation \cite{Weisberger:1966ip}. 
However, lattice calculations of all three form factors have shown that this relation is violated badly \cite{Rajan:2017lxk,Tsukamoto:2017fnm,Jang:2018lup,Ishikawa:2018rew,Liang:2018pis}. Ref.\ \cite{Rajan:2017lxk} concludes that lattice spacing artifacts cannot explain the large violation, leaving essentially excited-state effects as a natural explanation. Indeed, Ref.\ \cite{Bali:2018qus} argues that a large part of the violation may be due to strong excited-state contaminations in the 3-pt functions involving the pseudoscalar density and the time component of the axial vector current.

Here we present the results for the $N\pi$-state contamination in the pseudoscalar 3-pt function and lattice estimators for the pseudoscalar form factor $\GP(Q^2)$. The results are derived to LO in ChPT. Together with the analogous results for the axial form factors $\GA(Q^2)$ and $\GPt(Q^2)$ we explicitly calculate the dominant violations of the gGT relation by $\np$ excited states. 

As anticipated in \cite{Bali:2018qus} we find a sizeable $\np$-state contamination in lattice estimates for $\GP(Q^2)$. It leads to an underestimation for small $Q^2$, but to an overestimation for larger values. The size of the deviation depends on the source-sink separation $t$ in the pseudoscalar 3-pt function, and covers the range $-20$\% to +40\% for $t=2$ fm and momentum transfers below 0.25 GeV$^2$. The $Q^2$ dependence of the deviation leads to a distortion of the expected PPD behavior.  

The validity of the generalized Goldberger-Treiman relation is sometimes tested by a ratio $r_{\rm PCAC}$ \cite{Rajan:2017lxk,Bali:2018qus}. It involves all three form factors and the deviation from  the value 1 is a quantitative measure for a violation of the generalized Goldberger-Treiman relation. We find that the $\np$-state contamination in all three form factors results in $r_{\rm PCAC}< 1$, and the difference to 1 increases the smaller $Q^2$ is. In fact, comparing the LO ChPT results with lattice QCD data recently obtained by the PACS collaboration \cite{Ishikawa:2018rew} we find remarkable agreement even for source-sink separations as small as $t\approx 1.3$ fm. This supports the conclusion that the observed violation of the generalized Goldberger-Treiman relation has its origin in non-negligible excited-state contaminations.

The ChPT calculation of the $\np$ contamination in the axial form factors can be found in Ref.\ \cite{Bar:2018xyi}. It involves the computation of various Feynman diagrams stemming from the chiral expansion of the axial vector 3-pt function. In principle the same set of diagrams needs to be computed to obtain the  $\np$ contamination in the pseudoscalar form factor, with the axial vector current replaced by the pseudoscalar density. However, it is simpler to proceed differently. By construction, ChPT reproduces the chiral Ward identities of QCD, in particular the PCAC relation. Therefore, the PCAC relation can be used to directly obtain the $\np$ contamination in the pseudoscalar 3-pt function from the results for the axial vector 3-pt functions in \cite{Bar:2018xyi}. The same strategy is used in Ref.\ \cite{Bar:2012ce} in the computation of the 3-pion excited-state contribution to the QCD two-point functions of the axial-vector current and the pseudoscalar density. 

This paper relies heavily on the results in Ref.\ \cite{Bar:2018xyi}, and the reader is assumed to be familiar with this reference. The general ideas behind ChPT calculations of the  $\np$-state contamination in nucleon observables have been recently reviewed in \cite{Bar:2017kxh,Bar:2017gqh} and are not repeated here.

%
\section{Axial and pseudoscalar form factors of the nucleon}
%
\subsection{Basic definitions and results}
We follow Ref.\ \cite{Bar:2018xyi} and consider QCD with degenerate up and down quark masses. The spatial volume is assumed to be finite with extent $L$, and periodic boundary conditions are imposed for all spatial directions. The time extent is taken infinite, for simplicity, and we work in euclidean space time.

We are interested in the matrix element of the local iso-vector pseudoscalar density $P^a(x)$  between single-nucleon states of definite momenta and spin,
\begin{equation}\label{DefpseudoscalarFF}
m_q \langle N(p',s')|P^a(0)|N(p,s)\rangle = m_q \GP(Q^2)\bar{u}(p',s')\gamma_5 \frac{\sigma^a}{2}u(p,s)\,.
\end{equation}
Here $m_q$ denotes the mass of the up and down quarks.
The right hand side defines the pseudoscalar form factor $\GP(Q^2)$. $u(p)$ is an isodoublet Dirac spinor with momentum $p$ and spin $s$, and the four-momentum transfer $Q_{\mu}$ is given by 
\begin{equation}
Q_{\mu}=(i E_{N,\vec{p}^{\,\prime}} - i E_{N,\vec{p}},\vec{q})\qquad \vec{q} = \vec{p}^{\,\prime}-\vec{p}\,.
\end{equation}
In euclidean  (lattice) QCD the form factors are computed for space-like momentum transfers $Q^2>0$, with 
$Q^2=(\vec{p}^{\,\prime}-\vec{p})^2 - (E_{N,\vec{p}^{\,\prime}} - E_{N,\vec{p}})^2$ and $E_{N,\vec{p}}^2=\vec{p}^2+M_N^2$.

In analogy to the pseudoscalar density we also consider the analogous matrix element of the local iso-vector axial vector  current $A_{\mu}^a(x)$,
\begin{equation}\label{DefaxialFF}
\langle N(p',s')|A_{\mu}^a(0)|N(p,s)\rangle = \bar{u}(p',s')\left(\gamma_{\mu}\gamma_5 \GA(Q^2) - i \gamma_5\frac{Q_{\mu}}{2M_N}\GPt(Q^2)\right)\frac{\sigma^a}{2}u(p,s)\,.
\end{equation}
The right hand side shows the decomposition of the matrix element in two form factors, the axial form factor $\GA(Q^2)$ and the induced pseudoscalar form factor $\GPt(Q^2)$. 
The matrix elements of the axial vector current and the pseudoscalar density are not independent but related via the PCAC relation,
\begin{equation}\label{PCACrelation}
\partial_{\mu}A^a_{\mu}(x) = 2 m_q P^a(x)\,.
\end{equation}
Taking this relation between single nucleon (SN) states provides the gGT relation,
\begin{equation}\label{pcacformfactorlevel}
2M_N G_{\rm A}(Q^2) -\frac{Q^2}{2M_N} \GPt(Q^2) = 2 m_q \GP(Q^2)\,,
\end{equation}
between the three form factors.\footnote{Ref.\ \cite{Bali:2018qus} refers to it as the PCAC$_{\rm FF}$ relation.}

Considering \pref{pcacformfactorlevel} in the limit of vanishing momentum transfer and pion mass one can conclude that both  $\GPt(Q^2)$ and $ \GP(Q^2)$ are dominated by a pion pole for small $Q^2$. For $Q^2$ close to $-M_{\pi}^2$
one can derive the expressions\footnote{See appendix B of Ref.\ \cite{Sasaki:2007gw} for a quick derivation.} 
\begin{eqnarray}\label{ppdhypothesis}
\GPt^{\tiny \rm ppd}(Q^2)  &=& \frac{4M_N^2}{Q^2+M_{\pi}^2} G_{\rm A}(Q^2)\,,\label{ppd1}\\
2 m_q \GP^{\tiny \rm ppd}(Q^2) &=& \frac{2M_NM_{\pi}^2}{Q^2+M_{\pi}^2} G_{\rm A}(Q^2)\,,\label{ppd2}
\end{eqnarray}
for the form factors, which are called the PPD model results.

A simple ansatz for the axial form factor $\GA(Q^2)$, used commonly in fits to experimental data, is provided by the dipole approximation 
\begin{equation}\label{GAdip}
\GA^{\tiny \rm dip}(Q^2)= \frac{g_{\rm A}}{(1+Q^2/{ M}_{\rm A}^2)^2}\,,
\end{equation}
where  ${M}_{\rm A}$ is the axial dipole mass and $g_{\rm A}=\GA(0)$ the axial charge of the nucleon. It is a simple one parameter ansatz that reproduces the axial charge for $Q^2=0$ and the behavior $\GA(Q^2)\sim Q^{-4}$ for large momentum transfers expected from perturbation theory.
\subsection{Correlation functions}
The standard procedure to compute the form factors is based on evaluating various 2- and 3-point (pt) functions. Explicitly, the nucleon 2-pt function is given by
\begin{equation}\label{Def2ptfunc}
C_2(\vec{p},t)= \int d^3x \,e^{i\vec{p}\vec{x}}\, \Gamma_{\beta\alpha}\langle N_{\alpha}(\vec{x},t)\overline{N}_{\beta}(0,0)\rangle \,.
\end{equation}
$N,\overline{N}$ denote interpolating fields of the nucleon. We assume them to be given by the standard 3-quark operators \cite{Ioffe:1981kw,Espriu:1983hu} (either point like or smeared) that have been mapped to ChPT \cite{Wein:2011ix,Bar:2015zwa,Bar:2013ora}. The projector $\Gamma$ acts on spinor space and is given by
\begin{equation}\label{DefProjGamma}
\Gamma=\frac{1+\gamma_4}{4}(1+i \gamma_5\gamma_3)
\end{equation}
in terms of euclidean gamma matrices. Some lattice collaborations choose a different normalization for $\Gamma$, but this is irrelevant for the results in this paper.

In the following the nucleon 3-pt function is computed with the nucleon at the sink being at rest, i.e.\ $\vec{p}^{\,\prime}=0$. This implies $\vec{q}=-\vec{p}$ and  
\be
Q^2=2M_N (E_{N,\vec{q}} -M_N)
\ee
for the momentum transfer. In addition, we always choose the third isospin component of the axial vector current and the pseudoscalar density, i.e.\ $a=3$. Thus, the nucleon 3-pt functions we consider are given by
\begin{eqnarray}
C_{3,P^3}(\vec{q},t,t')&=&\int d^3x\int d^3y \,e^{i\vec{q}\vec{y}}\, \Gamma_{\beta\alpha}\langle N_{\alpha}(\vec{x},t) P^3(\vec{y},t')\overline{N}_{\beta}(0,0)\rangle\,,\label{C3ptP}\\
C_{3,A^3_{\mu}}(\vec{q},t,t')&=&\int d^3x\int d^3y \,e^{i\vec{q}\vec{y}}\, \Gamma_{\beta\alpha}\langle N_{\alpha}(\vec{x},t) A_{\mu}^3(\vec{y},t')\overline{N}_{\beta}(0,0)\rangle\,.\label{C3ptAmu}
\end{eqnarray}
The euclidean times $t$ and $t'$ denote the source-sink separation and the operator insertion time, respectively.
With the 2-pt and 3-pt functions we define the generalized ratios
\begin{eqnarray}\label{DefRatio}
R_{\mu}(\vec{q},t,t')& =&\frac{C_{3,X_{\mu}}(\vec{q},t,t')}{C_2(0,t)}\sqrt{\frac{C_2(\vec{q},t-t')}{C_2(0,t-t')}\frac{C_2(\vec{0},t)}{C_2(\vec{q},t)}\frac{C_2(\vec{0},t')}{C_2(\vec{q},t')}}\,,\quad \mu=1,\ldots 4,P\,.
\end{eqnarray}
For $\mu=1,\ldots 4$ the ratio involves the axial vector current 3-pt function \pref{C3ptAmu}. As a short hand notation we also allow for $\mu=P$, referring to the case with the 3-pt function \pref{C3ptP}. 
The ratios are defined in such a way that, in the asymptotic limit $t,t', t-t'\rightarrow \infty$, they converge to constant asymptotic values  $\Pi_{{\mu}}(\vec{q})$,
\begin{equation}
R_{\mu}(\vec{q},t,t') \rightarrow\Pi_{{\mu}}(\vec{q})\,.
\end{equation}
These are related to the form factors according to
\begin{eqnarray}
\Pi_{{k}}(\vec{q})& =& \frac{i}{\sqrt{2E_{N,\vec{q}}(M_N+ E_{N,\vec{q}})}}\left( (M_N+E_{N,\vec{q}})\GA(Q^2) \delta_{3k}-\frac{\GPt(Q^2)}{2M_N} q_3q_k\right),\label{AsympValueR33}\\
\Pi_4(\vec{q})&=& \frac{q_3}{\sqrt{2E_{N,\vec{q}}(M_N+ E_{N,\vec{q}})}}\left(\GA(Q^2)+\frac{M_N-E_{N,\vec{q}}}{2M_N}\GPt(Q^2)\right)\,,\label{AsympValueR30}\\
\Pi_{\rm P}(\vec{q}) &=&  \frac{q_3}{\sqrt{2E_{N,\vec{q}}(M_N+ E_{N,\vec{q}})}}\,\GP(Q^2)\,.\label{AsympValueP}
\end{eqnarray}
The PCAC relation in eq.\ \pref{PCACrelation} implies the constraint
\begin{equation}\label{pcacrelation3pt}
2m_q C_{3,P^3}(\vec{q},t,t') = \partial_{t'} C_{3,A^3_{4}}(\vec{q},t,t') - i \sum_{k=1}^3 q_k C_{3,A^3_{k}}(\vec{q},t,t')
\end{equation}
between the various 3-pt functions. Multiplying the 2-pt function contribution to form the ratios \pref{DefRatio} we obtain  
\begin{equation}\label{pcacratios}
2m_q R_{\rm P}(\vec{q},t,t') = R^{'}_{4}(\vec{q},t,t') - i \sum_{k=1}^3 q_k R_{k}(\vec{q},t,t')\,,
\end{equation}
where $R^{'}_{4}(\vec{q},t,t')$ denotes the ratio involving the time derivative $ \partial_{t'} C_{3,A^3_{4}}(\vec{q},t,t') $. Eq.\ \pref{pcacratios} is the PCAC relation on the level of the ratios. Taking the times $t,t'$ both to infinity it reduces to
\begin{equation}\label{RelPipPI0}
2m_q\Pi_{\rm P}(\vec{q}) = 2M_N \Pi_{\rm 4}(\vec{q})\,,
\end{equation}
i.e.\ the SN contribution of the pseudoscalar ratio is directly proportional to the one of the time component of the axial vector current. Together with eqs.\ \pref{AsympValueR30} and \pref{AsympValueP} we immediately reproduce the gGT relation \pref{pcacformfactorlevel}.
%
\section{Excited state analysis }
%
\subsection{Preliminaries}

In principle the form factors are obtained from the asymptotic values $\Pi_{\mu}(\vec{q})$ of the ratios. For example, the pseudoscalar form factor $\GP(Q^2)$ is directly proportional to $\Pi_{\rm P}(\vec{q})$. The proportionality factor is a simple kinematical factor that is easily computed and removed from $\Pi_{\rm P}(\vec{q})$. The two axial form factors $\GA(Q^2)$ and $\GPt(Q^2)$ are computed analogously, although in general one has to solve a linear system to extract the two form factors from two independent asymptotic values.\footnote{In lattice calculations one often measures more than two asymptotic values and constructs an overdetermined linear system for the two unknown form factors. This is subsequently solved by minimizing a suitably defined least-squares function \cite{Capitani:2017qpc,Alexandrou:2017hac}.}

In practice one only has access to the ratios $R_{\mu}(\vec{q},t,t')$ at time separations $t,t'$ that are far from being asymptotically large. In that case the correlation functions and the ratios not only contain the contribution of the lowest lying SN state, but also of excited states with the same quantum numbers as the nucleon. This excited-state contamination also enters the calculation of the form factors. Instead of the true form factors one is interested in one obtains {\em effective} form factors including an excited-state contamination. The effective form factors are expected to be of the form\footnote{For brevity we introduce the notation $G_{\tilde{\rm P}}=\tilde{G}_{\rm P}$.} 
\begin{eqnarray}
G^{\rm eff}_{\rm X}(Q^2,t,t')\, = \,G_{\rm X}(Q^2)\bigg[ 1 + \DG_{\rm X}(Q^2,t,t')\bigg],\quad X\,=\,A,P,\tilde{P}\,.\label{EffFF}
\end{eqnarray}
The excited-state contribution $\DG_{\rm X}(Q^2,t,t')$ vanishes for $t,t',t-t'\rightarrow \infty$. 

For pion masses as small as in Nature one can expect two-particle $N\pi$ states to cause the dominant excited-state contamination for large but finite time separations. This expectation rests on the naive observation that the energy gaps between the $N\pi$ states and the SN ground state are smaller than those one expects from true resonance states like the Roper resonance. Note that this not only requires small pion masses but also sufficiently large volumes such that the discrete spatial momenta imply small energies for the lowest-lying $N\pi$ states. Volumes with $M_{\pi}L\simeq 4$, often used in lattice simulations, already fulfill this criterion \cite{Bar:2017kxh}.

In this section we derive formulae that capture the $N\pi$-state contamination in the 2-pt and 3-pt functions, the ratio $R_{\mu}$ and eventually in the effective form factors. In these expression the $N\pi$-state contamination is parameterized in terms of coefficients stemming from ratios of various matrix elements with $N\pi$ states as initial and/or final states. In the next subsection ChPT will be used to compute these coefficients perturbatively.

\subsection{$\np$ states in the correlation functions}
 Performing the standard spectral decomposition in $C_2(\vec{q},t)$ defined in eq.\ \pref{Def2ptfunc}, the 2-pt function is a sum of various contributions,
\begin{eqnarray}\label{2ptDecomp}
C_2(\vec{q},t) & = & C^N_2(\vec{q},t) + C^{\np}_2(\vec{q},t)+\ldots\,.
\end{eqnarray}
The first two terms on the right hand side refer to the SN and the $N\pi$ contributions. The ellipsis refers to omitted contributions which we assume to be small in the following. The SN contribution is given by
\begin{equation}\label{SNcontr}
C^N_2(\vec{q},t)=\frac{1}{2E_{N,\vec{q}}}\;|\langle 0|N(0)|N(-\vec{q})\rangle|^2 e^{-E_{N,\vec{q}}\, | t |} \,.
\end{equation}
Here $|N(-\vec{q})\rangle$ denotes the state for a moving nucleon with momentum $-\vec{q}$. The interpolating field $N(0)$ also excites $N\pi$  states with the same quantum numbers as the nucleon, thus we obtain the non-vanishing $N\pi$ contribution 
\begin{eqnarray}\label{Npicontr}
C^{\np}_2(t)&=&\frac{1}{L^3}\;\sum_{\vec{k}}\frac{1}{4E_{N,\vec{r}} \Epik}\,
|\langle 0|N(0)|N(\vec{r}) \pi(\vec{k})\rangle|^2 e^{-E_{\rm tot}|t|}\,.
\end{eqnarray}
The sum runs over all pion momenta $\vec{k}$ that are compatible with the periodic boundary conditions, and the nucleon momentum is fixed to $\vec{r}=-\vec{q}-\vec{k}$. $E_{\rm tot}$ is the total energy of the $N\pi$ state. For weakly interacting pions $E_{\rm tot}$ equals approximately the sum $E_{N,\vec{r}}+ \Epik$ of the individual nucleon and pion energies. 

Since the leading SN contribution is nonzero we can rewrite eq.\ \pref{2ptDecomp} as
\begin{eqnarray}\label{2ptDecomp2}
C_2(\vec{q},t) & =& C^N_2(\vec{q},t)\left\{1+ \sum_{\vec{k}} d(\vec{q},\vec{k})e^{-\Delta E(\vec{q},\vec{k}) t}\right\}\,.\label{DefC2Npcontr}
\end{eqnarray}
The coefficient $d(\vec{q},\vec{k})$ is essentially the ratio of the matrix elements in eqs.\ \pref{Npicontr} and \pref{SNcontr}, and the energy gap $\Delta E(\vec{q},\vec{k})$ reads
\begin{equation}\label{Egap2pt}
\Delta E(\vec{q},\vec{k}) = E_{\pi,\vec{k}} + E_{N,\vec{q}+\vec{k}} - \ENq\,.
\end{equation}
As mentioned before, we have ignored the nucleon-pion interaction energy. Computing the 2-pt function in ChPT to LO one recovers the result  \pref{Egap2pt} for the energy gap \cite{Bar:2018xyi}. Deviations due to the nucleon-pion interaction will appear at higher order in the chiral expansion.

The 2-pt function enters the generalized ratio $R_{\mu}(\vec{q},t,t')$. Introducing the short hand notation $\sqrt{\Pi C_2}$ for the square root expression in \pref{DefRatio} and expanding in powers of small quantities we obtain
\begin{equation}\label{ratio2ptGeneric}
\frac{1}{C_2(0,t)}\sqrt{\Pi C_2} = \frac{1}{C^N_2(0,t)}\sqrt{\Pi C^N_2} \left\{ 1+ \frac{1}{2} Y(\vec{q},t,t')\right\}\,,
\end{equation}
where the function $Y(\vec{q},t,t')$ contains the $\np$-state contribution,
\begin{eqnarray}
Y(\vec{q},t,t')&=&\sum_{\vec{k}}\Big(d(\vec{q},\vec{k})  \left\{ e^{-\Delta E(\vec{q},\vec{k}) (t-t')} - e^{-\Delta E(\vec{q},\vec{k}) t'} - e^{-\Delta E(\vec{q},\vec{k}) t}\right\}\nn\\
& & \phantom{\sum_{\vec{k}}} - d(0,\vec{k})  \left\{ e^{-\Delta E(\vec{0},\vec{k}) (t-t')} - e^{-\Delta E(\vec{0},\vec{k}) t'} + e^{-\Delta E(\vec{0},\vec{k}) t}\right\}\Big)\,.\label{DefY}
\end{eqnarray}

The excited-state analysis of the 3-pt function is analogous. Performing again the spectral decomposition we find, in analogy to \pref{2ptDecomp}, the result ($\mu=1,\ldots,4,{P}$)
\begin{eqnarray}
C_{3,\mu}(\vec{q},t,t') & = & C^N_{3,\mu}(\vec{q},t,t') + C^{\np}_{3,\mu}(\vec{q},t,t')+\ldots\,,\label{C3muspecDecomp}\\
& =& C^N_{3,\mu}(\vec{q},t,t')\bigg(1+ Z_{\mu}(\vec{q},t,t')\bigg)\label{DefC3Npcontr}\,.
\end{eqnarray}
As before we ignore all but the SN and the $N\pi$ contribution in the following. Thus, $Z_{\mu}$ denotes the ratio $C^{\np}_{3,\mu}(\vec{q},t,t')/C^N_{3,\mu}(\vec{q},t,t')$. Forming this ratio we assume and only consider the cases where the SN contribution is non-vanishing, which puts a constraint on the possible momenta $\vec{q}$ and the index $\mu$. 

With the assumed kinematical setup the generic form for $Z_{\mu}(\vec{q},t,t')$ is found as
\begin{eqnarray}
Z_{\mu}(\vec{q},t,t') & = & a_{\mu}(\vec{q}) e^{-\Delta E(0,-\vec{q}) (t-t')}+ \tilde{a}_{\mu}(\vec{q}) e^{-\Delta E(\vec{q},-\vec{q})t'} \nn \\[2ex]
& & + \sum_{\vec{k}} b_{\mu}(\vec{q},\vec{k}) e^{-\Delta E(0,\vec{k}) (t-t')}+\sum_{\vec{k}}\tilde{b}_{\mu}(\vec{q},\vec{k}) e^{-\Delta E(\vec{q},\vec{k}) t'} \nn\\
& & +  \sum_{\vec{k}} c_{\mu}(\vec{q},\vec{k}) e^{-\Delta E(0,\vec{k}) (t-t')}e^{-\Delta E(\vec{q},\vec{k}) t'}\,.\label{DefZmu}
\end{eqnarray}
The coefficients $a_{\mu}(\vec{q}), \tilde{a}_{\mu}(\vec{q}),b_{\mu}(\vec{q},\vec{k}) ,\tilde{b}_{\mu}(\vec{q},\vec{k}), c_{\mu}(\vec{q},\vec{k})$ in \pref{DefZmu} contain ratios of matrix elements involving the nucleon interpolating fields and either the axial vector current or pseudoscalar density. For example, the coefficient $b_{P}(\vec{q},\vec{k})$ contains the matrix element $\langle N\pi |P^a|N \rangle$ with the $N\pi$ state as the final state.  Similarly, $\tilde{b}_{P}(\vec{q},\vec{k})$ contains the matrix element with the $N\pi$ state as the initial state. Together the $b_{P}(\vec{q},\vec{k})$ and $\tilde{b}_{P}(\vec{q},\vec{k})$ contributions form the excited-to-ground-state contribution. Similarly, the $c_{P}(\vec{q},\vec{k})$ contribution is called the excited-to-excited-state contribution, since it involves the matrix elements with $N\pi$ states as initial and final states. An explanation for the presence of the $a_{P}(\vec{q})$ and $\tilde{a}_{P}(\vec{q})$ contribution, which multiply the same exponentials as the $b_{P}(\vec{q},-\vec{q})$ and $\tilde{b}_{P}(\vec{q},-\vec{q})$ contribution, will be given in the next section.

Taking the product of \pref{DefC3Npcontr} and \pref{ratio2ptGeneric} we obtain the total result for the $N\pi$ contamination in the generalized ratios,
\begin{eqnarray}
R_{\mu}(\vec{q},t,t')&=& \Pi_{\mu}(\vec{q}) \Bigg(1+ Z_{\mu}(\vec{q},t,t') + \frac{1}{2} Y(\vec{q},t,t')\Bigg)\,,\\
& \equiv & \Pi_{\mu}(\vec{q}) \Bigg(1+ X_{\mu}(\vec{q},t,t')\Bigg)\,,\label{NpiConttot}
\end{eqnarray}
with $\Pi_{\mu}(\vec{q})$ referring to the asymptotic values of the ratios introduced in \pref{AsympValueR33} -  \pref{AsympValueP}. The $N\pi$ contamination $X_{\mu}(\vec{q},t,t')$ vanishes exponentially as the time separations tend to infinity, so the ratios  correctly approach their asymptotic values. 

The pseudoscalar form factor is directly proportional to the asymptotic value $\Pi_{\rm P}(\vec{q})$. Therefore, comparing \pref{NpiConttot} with \pref{EffFF} we read off the simple relation $\DG_{\rm P}(Q^2,t,t')=X_{\rm P}(Q^2,t,t')$. In case of the axial form factors the relation between the $X_{\mu}(Q^2,t,t')$ and the $\DG_{{\rm A},{\tilde{\rm P}}}(Q^2,t,t')$ is slightly more involved and depends on the particular choice for the ratios one has made to extract the form factors. For details see Ref.\ \cite{Bar:2018xyi}.

\subsection{The PCAC relation and the $\np$ contribution  $Z_{\rm P}(\vec{q},t,t')$}

The  PCAC relation relates the 3-pt functions of the pseudoscalar density and the axial vector current. This relation not only holds for the SN contribution but for all contributions in the spectral decomposition. Consequently, the coefficient $a_{P}(\vec{q},\vec{k})$, for example, is related to and computable in terms of the coefficients $a_{\mu}(\vec{q},\vec{k})$. The same holds for all the other coefficients in \pref{DefZmu}.

To derive these relations we use the spectral decomposition \pref{C3muspecDecomp} in the PCAC relation in eq.\ \pref{pcacrelation3pt}. Since the SN contribution satisfies $2m_q C_{3,P}^N(\vec{q},t,t') = 2M_N C_{3,4}^N(\vec{q},t,t')$ we obtain
\begin{eqnarray}
2m_q C_{3,P}(\vec{q},t,t') &=& 2 M_N C_{3,4}^N (\vec{q},t,t') +\partial_{t'} C^{N\pi}_{3,4}(\vec{q},t,t') - i q_k C^{N\pi}_{3,k}(\vec{q},t,t')\,.
\end{eqnarray}
Here and in the following a sum over the spatial index $k=1,2,3$ is implied on the right hand side.
Provided  $C_{3,4}^N(\vec{q},t,t')\neq0$ this is easily brought into the  form 
\begin{eqnarray}\label{intermed1}
2m_q C_{3,P}(\vec{q},t,t') &=& 2 M_N C_{3,4}^N(\vec{q},t,t')\Big(1+ Z_4^{\prime}(\vec{q},t,t')+ \alpha_k(\vec{q})Z_k(\vec{q},t,t')   \Big)\,.
\end{eqnarray}
The newly introduced $\alpha_k$ are the short hand notation for the combination
\begin{equation}\label{Defalphak}
\alpha_k(\vec{q}) =  -i  \frac{C_{3,k}^N(\vec{q},t,t')}{C_{3,4}^N(\vec{q},t,t')} \frac{q_k}{2M_N}\,.
\end{equation}
Note that the time dependence of the 3-pt functions cancels in the ratio on the right hand side, thus $\alpha_k$ is a constant for fixed momentum $\vec{q}$.
The remaining term involving the time derivative, 
\begin{eqnarray}
Z_4^{\prime}(\vec{q},t,t')& \equiv&  \frac{\partial_{t'} C^{N\pi}_{3,4}(\vec{q},t,t')}{2M_N C_{3,4}^N(\vec{q},t,t')}\,,
\end{eqnarray}
has the same form as the original $Z_4(\vec{q},t,t')$, but with primed coefficients 
$a^{\prime}_{4}(\vec{q}), \tilde{a}^{\prime}_{4}(\vec{q}),b^{\prime}_{4}(\vec{q},\vec{k}) ,\tilde{b}^{\prime}_{4}(\vec{q},\vec{k}), c^{\prime}_{4}(\vec{q},\vec{k})$.
The primes serve as a reminder that the coefficients involve additional factors stemming from the time derivative $\partial_{t'}$ of the exponentials in $C^{N\pi}_{3,4}(\vec{q},t,t')$:
\begin{eqnarray}
a^{\prime}_{4}(\vec{q}) &=&  \frac{E_{\pi,\vec{q}}}{2M_N} \,\,a_{4}(\vec{q}),\label{apP0}\\
\tilde{a}^{\prime}_{4}(\vec{q}) &=& -\frac{E_{\pi,\vec{q}}}{2M_N} \,\,\tilde{a}_{4}(\vec{q}),\label{atpP0}\\
b^{\prime}_{4}(\vec{q},\vec{k}) & =& \frac{E_{\pi,\vec{k}} +E_{N,\vec{k}} - E_{N,\vec{q}}}{2M_N} \,\,b_{4}(\vec{q},\vec{k}),\\
\tilde{b}^{\prime}_{4}(\vec{q},\vec{k}) &=& -\frac{E_{\pi,\vec{k}} -(E_{N,\vec{k}+\vec{q}} - E_{N,\vec{q}}) + (E_{N,\vec{q}} - M_N)}{2M_N} \,\, \tilde{b}_{4}(\vec{q},\vec{k}),\\
c_{4}^{\prime}(\vec{q},\vec{k}) & =& - \frac{E_{N,\vec{k}+\vec{q}} - E_{N,\vec{k}}}{2M_N}\,\,c_{4}(\vec{q},\vec{k}).\label{cpP0}
\end{eqnarray}
Putting everything together we obtain the following coefficients for the $N\pi$ contribution in $Z_{\rm P}(\vec{q},t,t')$ 
\begin{eqnarray}
a_{P}(\vec{q}) &=&a^{\prime}_{4}(\vec{q}) + \alpha_k a_{k}(\vec{q}),\label{aP0}\\
\tilde{a}_{P}(\vec{q}) &=&\tilde{a}^{\prime}_{4}(\vec{q}) + \alpha_k \tilde{a}_{k}(\vec{q}),\label{aPt0}\\
b_{P}(\vec{q},\vec{k}) &=&b^{\prime}_{4}(\vec{q},\vec{k}) + \alpha_k b_{k}(\vec{q},\vec{k}),\\
\tilde{b}_{P}(\vec{q},\vec{k}) &=&\tilde{b}^{\prime}_{4}(\vec{q},\vec{k}) + \alpha_k \tilde{b}_{k}(\vec{q},\vec{k}),\\
c_{P}(\vec{q},\vec{k}) &=&c^{\prime}_{4}(\vec{q},\vec{k}) + \alpha_k c_{k}(\vec{q},\vec{k}).\label{cP0}
\end{eqnarray}

\subsection{ChPT results for the coefficients}

The coefficients introduced in the previous subsection can be perturbatively computed in ChPT.\footnote{The first account for ChPT calculations of $\np$-state contributions is given in \cite{Tiburzi:2009zp}.} 
This has been done in Ref.\ \cite{Bar:2018xyi} for the coefficients with $\mu=1,\ldots,4$, i.e.\ for the correlation functions and ratios involving the axial vector current. To this end twelve 1-loop and three tree-level Feynman diagrams were computed to obtain the leading $\np$-state contribution to the correlation functions.\footnote{See fig.\ 3 in Ref.\ \cite{Bar:2018xyi} for the diagrams.} In principle, the same diagrams with $A^a_{\mu}$ replaced by $P^a$ need to be computed to obtain the $\np$-state contribution for the pseudoscalar correlation functions. Alternatively, since the PCAC relation is satisfied in ChPT, we can use the results \pref{aP0} - \pref{cP0} to get the pseudoscalar coefficients from the axial vector ones.

The calculations in \cite{Bar:2018xyi} were performed in the  covariant formulation of Baryon ChPT \cite{Bar:2018xyi}. The expressions for the coefficients are fairly cumbersome in the full covariant form. They simplify significantly if we perform the non-relativistic (NR) expansion of the nucleon energy,
\begin{equation}\label{NRexpansion}
\ENq = M_N+\frac{\vec{q}^{\,2}}{2M_N} \,,
\end{equation}
and keep the first two terms only. For practical uses this approximation is expected to be sufficient. For example, the NR expansion for the coefficients $a_{k}(\vec{q})$ reads
\begin{equation}
a_{k}(\vec{q}) = a^{\infty}_{k}(\vec{q}) +\frac{E_{\pi,\vec{q}}}{M_N} a^{\rm corr}_{k}(\vec{q})\,,
\end{equation}
and the results for $a^{\infty}_{k}(\vec{q}),a^{\rm corr}_{k}(\vec{q})$ are given in \cite{Bar:2018xyi}, eqs.\ (4.14) -- (4.17). Analogous expressions hold for the other coefficients.\footnote{The NR expansion of the coefficients with $\mu=4$ is slightly different, see \cite{Bar:2018xyi}.}

To compute the coefficients with $\mu=P$ according to \pref{aP0} - \pref{cP0} we need $\alpha_k(\vec{q})$ defined in \pref{Defalphak}. The LO ChPT results for $C_{3,k}^N(\vec{q},t,t')$ and $C_{3,4}^N(\vec{q},t,t')$ are given in \cite{Bar:2018xyi}, eqs.\ (4.2) and (4.3), respectively. Taking the ratio and performing the NR expansion we obtain
\begin{eqnarray}
\alpha_{k} & = &-\frac{q_k^2}{M_{\pi}^2}, \quad k=1,2\,, \qquad\quad  \alpha_{3} \, =\,\frac{\Epiq^2 -q_3^2}{M_{\pi}^2}\,.
\end{eqnarray}
The explicit expressions for the coefficients entering \pref{aP0} - \pref{cP0} are given in Ref.\ \cite{Bar:2018xyi}, section IV E.
In principle there is no need to write down the explicit results one obtains from \pref{aP0} - \pref{cP0}, in particular since the full expressions are quite cumbersome and not very illuminating. The leading results in the NR expansion, however, assume a compact form and may be useful for lattice practitioners in their analysis of lattice data, so we quote these results here.

As mentioned before, the calculation in Ref.\ \cite{Bar:2018xyi} involves various 1-loop and tree-level diagrams. It turned out to be convenient to introduce separate coefficients for the contributions originating in either 1-loop or tree diagrams. The coefficients $a_{\mu}(\vec{q})$ and $\tilde{a}_{\mu}(\vec{q})$ are associated to the latter. With \pref{aP0} and \pref{aPt0} we obtain
\begin{eqnarray}
a^{\infty}_{P}(\vec{q}) &=&\tilde{a}^{\infty}_{P}(\vec{q})\,\,=\,-\frac{1}{2}.\label{aInftyP0}
\end{eqnarray}
The remaining coefficients capture the loop diagram contribution. Following \cite{Bar:2018xyi} they are split into a universal part (containing the anticipated $1/L^3$ factor of a two-particle state in a finite spatial volume) and a ``reduced coefficient'',
\begin{equation}\label{DefRedCoeff}
b_P(\vec{q},\vec{k}) = \frac{1}{8 (fL)^2 E_{\pi,\vec{k}}L} B_P(\vec{q},\vec{k}).
\end{equation}
The NR expansion for the reduced coefficient is as before, 
\begin{equation}
B_{P}(\vec{q},\vec{k}) =B_{P}^{\infty}(\vec{q},\vec{k}) + \frac{E_{\pi,\vec{k}}}{M_N}B_{P}^{\rm corr}(\vec{q},\vec{k})\,,
\end{equation}
and analogous formulae hold for $\tilde{B}_{P}, C_{P}$. For the leading O(1) coefficients we obtain the following results (for brevity we introduce $kq\equiv k_{\mu}q_{\mu}$):
\begin{eqnarray}
B^{\infty}_{P}(\vec{q},\vec{k}) &=& 2g_A^2\frac{\Epiq^2}{M_{\pi}^2}\left(\frac{k^2}{\Epip^2} +\frac{k_3}{q_3}\frac{kq}{\Epip^2} \right)- 4\frac{\Epiq^2}{M_{\pi}^2}\frac{k_3}{q_3}\,,\label{Binfty}\\
\tilde{B}^{\infty}_{P}(\vec{q},\vec{k}) &=&2g_A^2\frac{\Epiq^2}{M_{\pi}^2}\left(\frac{k^2}{\Epip^2} +\frac{k_3}{q_3}\frac{kq}{\Epip^2} \right)+ 4\frac{\Epiq^2}{M_{\pi}^2}\frac{k_3}{q_3}\,,\\
C^{\infty}_{P}(\vec{q},\vec{k}) &=& g_A^2\frac{\Epiq^2}{M_{\pi}^2}\left(\frac{k^2}{\Epip^2} -2\frac{k_3}{q_3}\frac{kq}{\Epip^2} \right)\,.\label{Cinfty}
\end{eqnarray}
The $\np$ contribution to the ratios involves a sum over all discrete pion momenta allowed by the periodic boundary conditions. Some terms essentially average away when this sum is performed, for instance the last term proportional to $-4 k_3/q_3$ in \pref{Binfty}. However, the other terms contribute for all possible $\vec{k}$, leading to a non-vanishing contribution of the loop diagrams to the total $\np$-state contribution. 
%
\section{Impact on lattice calculations}
%
\subsection{Preliminaries}

To LO in ChPT the $N\pi$ contribution to the ratio $R_{\mu}$ and the effective form factors depends on a few low energy coefficients (LECs) only, and these are known rather precisely from experiment. Assuming these values in the ChPT results we obtain estimates for the expected impact of the $N\pi$ contribution  in lattice QCD simulations. 
The rationale for this application of the ChPT results is the same as for the axial form factors presented in Ref.\ \cite{Bar:2018xyi}. The reader is refered to Sect.\ V in this reference for details, here we merely summarize the values for the various input parameters that need to be fixed for the analysis.

Two necessary LECs are the chiral limit values of the pion decay constant and the axial charge.  To LO it is consistent to use the experimental values for these LECs and we set  $g_A=1.27$ and $f=f_{\pi}= 93$ MeV \cite{Tanabashi:2018oca}. We ignore the errors in these values since they are too small to be significant for the LO estimates. Two more LECs are associated with the pion and nucleon mass. Since we are mainly interested in the $N\pi$ contribution in physical point simulations we fix the pion and nucleon masses to their (approximate) physical values $M_{\pi}=140$ MeV and $M_N=940$ MeV. 

The spatial volume determines the accessible spatial momenta. In practice it is fixed by the the lattice spacing and the number of lattice points in the spatial directions. Typical values in recent lattice calculations cover a range $M_{\pi}L \sim 3$ to 6, and we will assume such values in the following. 
Imposing periodic boundary conditions the spatial momentum transfer can assume the values
$\vec{q}_n=(2\pi/L)\vec{n}_q$
with the vector $\vec{n}_q$ having integer valued components. 
These momenta imply the discrete values
\begin{equation}\label{DiscreteQ2}
Q^2_n= q_n^2\left(1-\frac{q_n^2}{4M_N^2}\right)
\end{equation}
for the 4-momentum transfer if we perform the NR expansion \pref{NRexpansion}.

ChPT is an expansion in the small pion mass and in small pion momenta. Therefore, we need to select an upper bound on the pion momentum in the $N\pi$ state. Following Refs.\ \cite{Bar:2016uoj,Bar:2016jof} we choose $|\vec{k}_n|\lesssim k_{\rm max}$ with $k_{\rm max}/\Lambda_{\chi}= 0.45$, where the chiral scale $\Lambda_{\chi}$ is equal to $4\pi f_{\pi}$. $N\pi$ states with pions satisfying this bound are called {\em low-momentum $N\pi$ states} in the following. For these we expect the LO ChPT results to work reasonably well. States with pion momenta larger than this bound are called {\em high-momentum $N\pi$ states}. These too contribute to the excited-state contamination. However, choosing all euclidean time separations sufficiently large the contribution of the high-momentum $N\pi$ states can be made small and negligible. The results in Refs.\ \cite{Bar:2016uoj,Bar:2016jof} suggest that at least a 1 fm separation between the operator and both source and sink is necessary for a sufficient suppression. This corresponds to source-sink separations of 2 fm or larger in the 3-pt functions. 

Note that an upper bound $|\vec{k}_n|\lesssim k_{\rm max}$ translates into a number $n_{k, {\rm max}}$ that depends on the spatial volume, i.e.\ on $M_{\pi}L$. The larger the volume the more discrete momenta satisfy the bound. Table \ref{table:npvalues} lists $n_{k, {\rm max}}$ for the volumes considered in this paper.\footnote{See Ref.\ \cite{Bar:2017kxh} for other upper momentum bounds.}

\begin{table}[bt]
\begin{center}
\begin{tabular}{l|c|c|c|c|}
\multirow{2}{*}{$\frac{k_{{\rm max}}}{\Lambda_{\chi}}$}& \multicolumn{4}{c|}{$n_{k,{\rm max}}$ }  \\ 
& $M_{\pi}L=3$ & $M_{\pi}L=4$ & $M_{\pi}L=5$ & $M_{\pi}L=6$  \\  \hline
0.45& 3 & 5 & 8&12 
\end{tabular}
\caption{{\label{table:npvalues}} $n_{k,{\rm max}}$ as a function of  $M_{\pi}L$ for $k_{{\rm max}}/\Lambda_{\chi}=0.45$.}
\end{center}
\end{table}

\subsection{Impact on the pseudoscalar form factor}\label{ssect:impactonPlatest}

\begin{figure}[t]
\begin{center}
\includegraphics[scale=1.0]{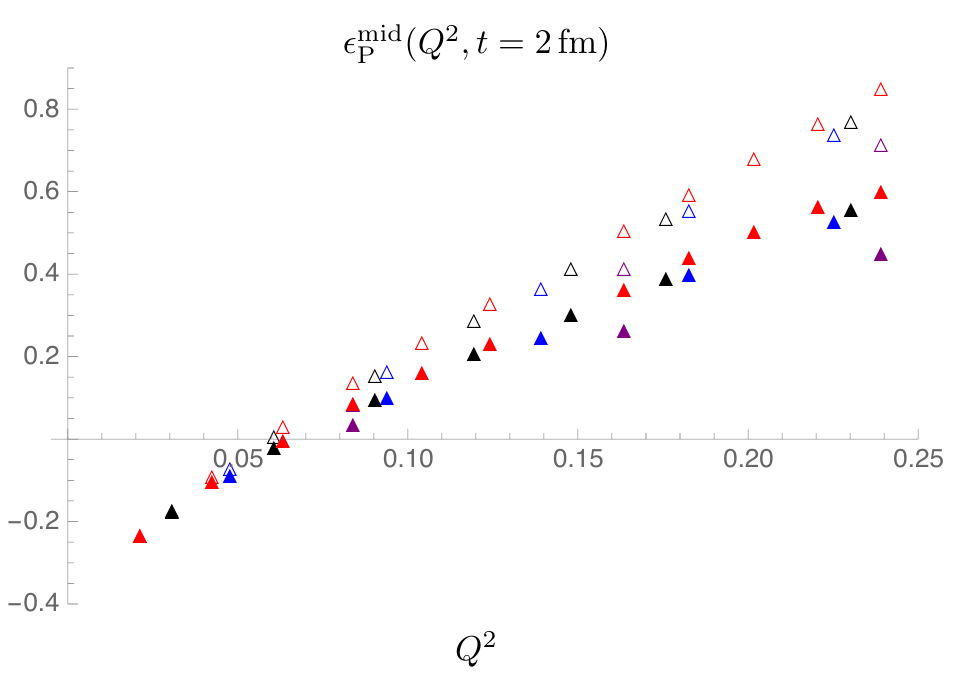}\\
\caption{\label{fig:epsP}The deviation $\epsP^{\rm mid}(Q^2,t)$  for $t=2$ fm as a function of $Q^2$ according \pref{DiscreteQ2} with $M_{\pi}L=3$ (purple), 4 (blue), 5 (black) and 6 (red). Open symbols correspond to the results with the leading NR limit values for the coefficients, filled symbols include the $1/M_N$ corrections.
}
\end{center}
\end{figure}

The effective form factors $G^{\rm eff}_{\rm X}(Q^2,t,t')$ depend on the source-sink separation $t$ and the operator insertion time $t'$. For fixed $t$ we introduce the  plateau estimates that, as a function of $t'$, minimize the deviation from the true form factors. The results of the last section imply $\DG_{\rm P}(Q^2,t,t')>0$, thus we define the plateau estimate 
\begin{eqnarray}
G_{\rm P}^{\rm plat}(Q^2,t)&\equiv \min\limits_{0<t'<t}G_{\rm P}^{\rm eff}(Q^2,t,t')\label{DefPlatEstGP}\,.
\end{eqnarray}
This is a function of the momentum transfer and $t$. Alternatively one can define a second estimator, the midpoint estimate
\begin{eqnarray}
G_{\rm P}^{\rm mid}(Q^2,t)&\equiv&G_{\rm P}^{\rm eff}(Q^2,t,t'=t/2)\,.\label{DefMidpointEstimateGP}
\end{eqnarray}
With the ChPT results for the $\np$ contribution the difference between the two estimators is very small, much smaller than the uncertainty one anticipates for the LO results.

As a measure for the $N\pi$-state contribution we introduce the relative deviation  from the true form factors,
\begin{eqnarray}
\epsilon^{\rm est}_{\rm P}(Q^2,t)\equiv \frac{G^{\rm est}_{\rm P}(Q^2,t)}{G_{\rm P}(Q^2)} -1\,,\label{DefEpsilons}
\end{eqnarray}
where ``est'' labels the plateau (plat) and mid-point (mid) estimators.

Figure \ref{fig:epsP} shows the results for $\epsP^{\rm mid}(Q^2,t)$. The source-sink separation was chosen as $t=2$ fm. The discrete $Q^2$ values stem from \pref{DiscreteQ2} with various different volumes satisfying $M_{\pi}L=3$ (purple), 4 (blue), 5 (black) and 6 (red). Open symbols correspond to the results with the leading NR limit values for the coefficients, see \pref{aInftyP0} -- \pref{Cinfty}, filled symbols include the $1/M_N$ corrections. For the largest momentum transfers displayed in fig.\ \ref{fig:epsP} the difference between the open and filled symbols is about 0.2. In the following we always include the $1/M_N$ corrections in the $\np$ contributions.

According to figure \ref{fig:epsP} the mid-point estimate  $\GP^{\rm mid}$ underestimates the physical form factor for $Q^2 \lesssim 0.06\,({\rm GeV})^2$, but overestimates for larger momentum transfers. Comparing the $M_{\pi}L=3$ and 6 results we do observe a noticeable finite volume (FV) effect that increases for larger $Q^2$. The results for $M_{\pi}\gtrsim 4$, however, fall essentially on a smooth line. 

\begin{figure}[t]
\begin{center}
\includegraphics[scale=1.0]{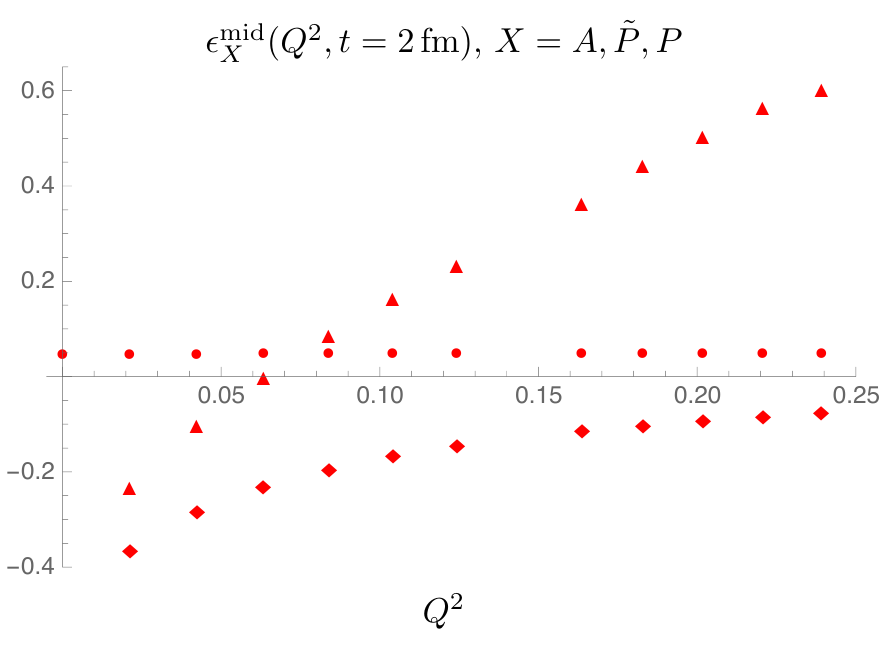}\\
\caption{\label{fig:epsAll}The relative deviations $\epsA^{\rm mid}(Q^2,t)$ (circles), $\epsPt^{\rm mid}(Q^2,t)$ (diamonds) and $\epsP^{\rm mid}(Q^2,t)$ (triangles) defined in \pref{DefEpsilons} as a function of $Q^2$ for $t=2$ fm. $Q^2$ values according to \pref{DiscreteQ2} with $M_{\pi}L=6$.}
\end{center}
\end{figure}

The analogous results for the axial form factor estimators are discussed in Ref.\ \cite{Bar:2018xyi}. Figure \ref{fig:epsAll} shows the relative deviations for all three form factors for $t=2$ and $M_{\pi}L=6$.
$\epsA^{\rm mid}(Q^2,t)$ is approximately $+0.05$ and essentially independent of the momentum transfer. $\epsPt^{\rm mid}(Q^2,t)$ shows a clear dependence on $Q^2$ and ranges  between $-0.4$ and $-0.05$ for the momentum transfers covered in the figure. While 
$\epsPt^{\rm mid}(Q^2,t)$ decreases for larger $Q^2$ the deviation $\epsP^{\rm mid}(Q^2,t)$ increases to values larger than 0.5 for $Q^2 \gtrsim 0.2\,({\rm GeV})^2$. Note that ChPT is expected to work better the smaller the momentum transfer is, i.e.\ higher order corrections are expected to become larger for larger $Q^2$.

\subsection{Impact on testing the gGT relation}

\begin{figure}[t]
\begin{center}
\includegraphics[scale=1.0]{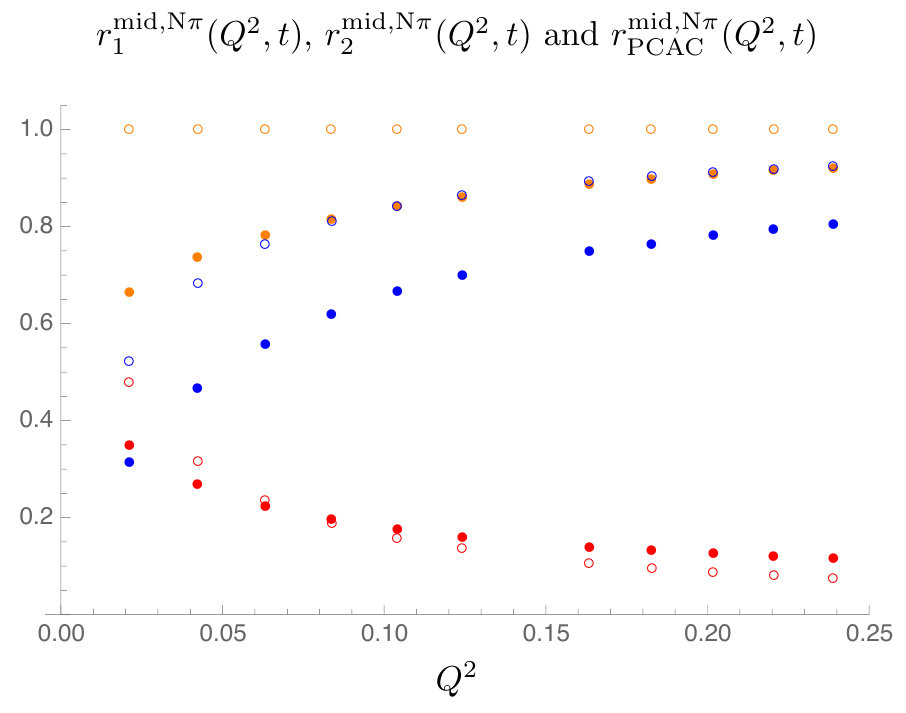}\\
\caption{\label{fig:r1r2rPCAC}The ratios $r^{\rm mid,\np}_{\rm 1}(Q^2,t)$ (blue), $r^{\rm mid,\np}_{\rm 2}(Q^2,t)$ (red) and $r^{\rm mid,\np}_{\rm PCAC}(Q^2,t)$ (orange) given in \pref{restNpiPCAC} - \pref{restNpi2} for the lowest discrete momenta corresponding to $M_{\pi}L=6$ at $t=2$ fm (filled symbols) and for infinite $t$ (open symbols). }
\end{center}
\end{figure}

As a measure to test the validity of the gGT relation one can introduce the ratios  \cite{Rajan:2017lxk,Bali:2018qus}
\begin{eqnarray}
r^{\rm est}_{\rm 1}(Q^2,t) & = &\frac{Q^2}{4M_N^2}\frac{\GPt^{\rm est}(Q^2,t)}{\GA^{\rm est}(Q^2,t)}\,,\\
 r^{\rm est}_{\rm 2}(Q^2,t) & = & \frac{2m_q}{2M_N}\frac{\GP^{\rm est}(Q^2,t)}{\GA^{\rm est}(Q^2,t)}\,,
\end{eqnarray}
and the sum 
\begin{eqnarray}\label{restNpiPCAC}
r^{\rm est}_{\rm PCAC}(Q^2,t)& =& r^{\rm est}_{\rm 1}(Q^2,t) + r^{\rm est}_{\rm 2}(Q^2,t)\,.
\end{eqnarray}
In the limit $t\rightarrow\infty$ the estimators converge to the SN form factors, and because of the gGT relation the ratio $r^{\rm est}_{\rm PCAC}(Q^2,t)$ assumes the constant value 1 in this limit.
For finite source-sink separations the excited state contribution in the  estimators cause a deviation from this value. With the results of the LO ChPT calculation the deviation due to $\np$ states is given by
\begin{eqnarray}
r^{{\rm est},\np}_{\rm 1}(Q^2,t) & = & \frac{Q^2}{Q^2+M_{\pi}^2}\left[\frac{1+\epsilon^{\rm est}_{\rm \tilde{P}}(Q^2,t)}{1+\epsilon^{\rm est}_{\rm A}(Q^2,t)}\right] \,,\label{restNpi1}\\
 r^{{\rm est},\np}_{\rm 2}(Q^2,t) & = & \frac{M_{\pi}^2}{Q^2+M_{\pi}^2}\left[\frac{1+\epsilon^{\rm est}_{\rm P}(Q^2,t)}{1+\epsilon^{\rm est}_{\rm A}(Q^2,t)}\right]\,.\label{restNpi2}
\end{eqnarray}
Figure \ref{fig:r1r2rPCAC} shows these two ratios and their sum for $M_{\pi}L=6$ at $t=2$ fm (filled symbols). The open symbols correspond to the limit $t\rightarrow \infty$  without the $\np$ contribution. The blue and red symbols show the results for the two ratios $r^{\rm mid}_1$ and $r^{\rm mid}_2$, respectively. The plot reflects what we have already observed in figure \ref{fig:epsP}: The $\np$ contribution in the induced pseudoscalar form factor underestimates the true form factor, thus the filled blue symbols are below the open ones over the entire $Q^2$ range. The smaller the momentum transfer the larger the deviation. The $\np$ contribution in the pseudoscalar form factor underestimates for small $Q^2$ but overestimates for large $Q^2$. Therefore, the filled red symbols are below the open ones for small $Q^2$, and above for large $Q^2$. The sum $r^{\rm mid}_{\rm PCAC}$ (orange symbols) is smaller than 1 for the momentum transfers considered, and the difference is larger for smaller $Q^2$.

Two features are worth emphasizing. Firstly, we have seen that the $\np$ contribution in the pseudoscalar case significantly overestimates for larger $Q^2$, according to fig.\ \ref{fig:epsP} about 50\% for $Q^2\approx 0.2\,({\rm GeV})^2$. Testing the gGT relation with the ratio $r^{\rm est}_{\rm PCAC}$ this sizable overestimation is largely suppressed because the ratio $r^{\rm est}_2$ contributes only about 10\% to $r^{\rm est}_{\rm PCAC}$  for  $Q^2\gtrsim 0.15\,{\rm (GeV)}^2$. Secondly, for the larger $Q^2$ values the deviations of $r^{\rm est}_1$ and $r^{\rm est}_2$ partially compensate if the sum is taken. Overall it is fair to say that the violation of $r^{\rm est}_{\rm PCAC}=1$ stems dominantly from the $\np$ contamination in the induced pseudoscalar form factor, at least for the momentum transfers displayed in fig.\ \ref{fig:r1r2rPCAC}.

\subsection{Impact on testing the PPD hypothesis}

As measures to test the validity of the PPD hypothesis one can introduce the two ratios,
\begin{eqnarray}
r^{\rm est}_{\rm 3}(Q^2,t)& \equiv &  \frac{Q^2 +M_{\pi}^2}{4M_N^2} \frac{\GPt^{\rm est}(Q^2,t)}{\GA^{\rm est}(Q^2,t)}\,, \\
r^{\rm est}_{\rm 4}(Q^2,t)& \equiv & \frac{Q^2 +M_{\pi}^2}{2M_NM_{\pi}^2} \frac{2m_q\GP^{\rm est}(Q^2,t)}{\GA^{\rm est}(Q^2,t)}\,.
\end{eqnarray}
In the limit $t\rightarrow\infty$ these ratios converge to 1 provided the PPD results \pref{ppd1} and \pref{ppd2} are valid. Deviations due to excited states are expected for finite source-sink separations. 
With the LO ChPT results for the $\np$ state contributions we obtain
\begin{eqnarray}
r^{{\rm est}, \np}_{\rm 3}(Q^2,t)& = & \frac{1+\epsilon^{\rm est}_{\rm \tilde{P}}(Q^2,t)}{1+\epsilon^{\rm est}_{\rm A}(Q^2,t)}\,, \label{restNpi3}\\
r^{{\rm est},\np}_{\rm 4}(Q^2,t)& = & \frac{1+\epsilon^{\rm est}_{\rm {P}}(Q^2,t)}{1+\epsilon^{\rm est}_{\rm A}(Q^2,t)}\,.\label{restNpi4}
\end{eqnarray}

\begin{figure}[t]
\begin{center}
\includegraphics[scale=1.0]{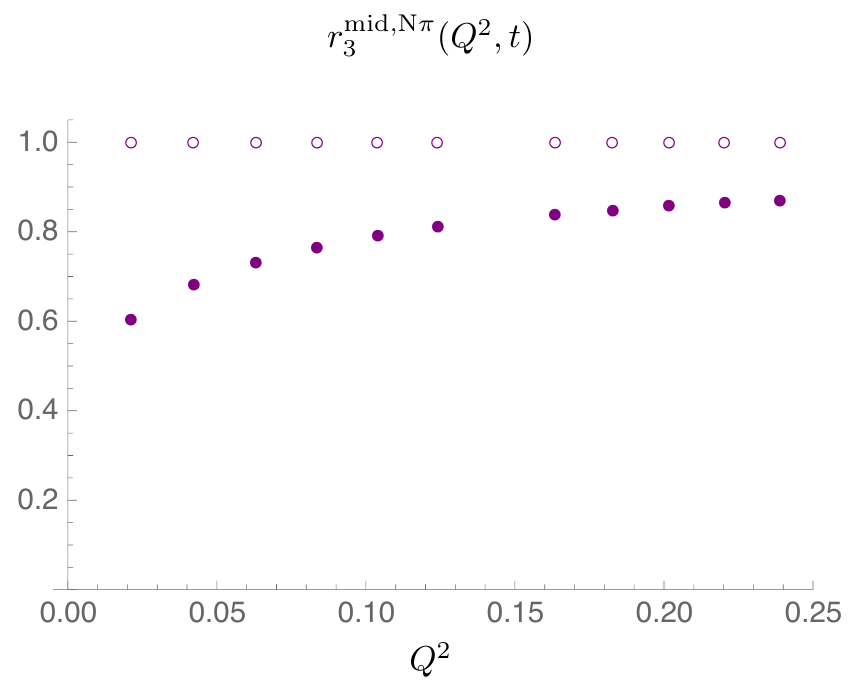}\\
\caption{\label{fig:r3}The ratio  $r^{\rm mid,\np}_{\rm 3}(Q^2,t)$, given in \pref{restNpi3}, for the lowest discrete momenta corresponding to $M_{\pi}L=6$ at $t=2$ fm (filled symbols) and for infinite $t$ (open symbols).}
\end{center}
\end{figure}
Figure \ref{fig:r3}  shows the ratio $r^{{\rm est}, \np}_{\rm 3}(Q^2,t)$, as before for $t=2$ fm and $M_{\pi}L=6$. The values are between 0.6 and 0.8 for the momentum transfers considered, i.e.\ significantly below 1, the SN result (open symbols). As mentioned before, the dominant reason for this difference is the underestimation of the induced pseudoscalar form factor, which is more prominent for small momentum transfers.

The $Q^2$ dependence of $r^{{\rm mid}, \np}_{\rm 3}$ resembles the one of $r^{{\rm mid}, \np}_{\rm PCAC}$. This is not surprising since $r^{{\rm mid}, \np}_{\rm PCAC}$ is, as discussed before, dominated by the $\np$ contamination in the induced pseudoscalar form factor

\begin{figure}[t]
\begin{center}
\includegraphics[scale=1.0]{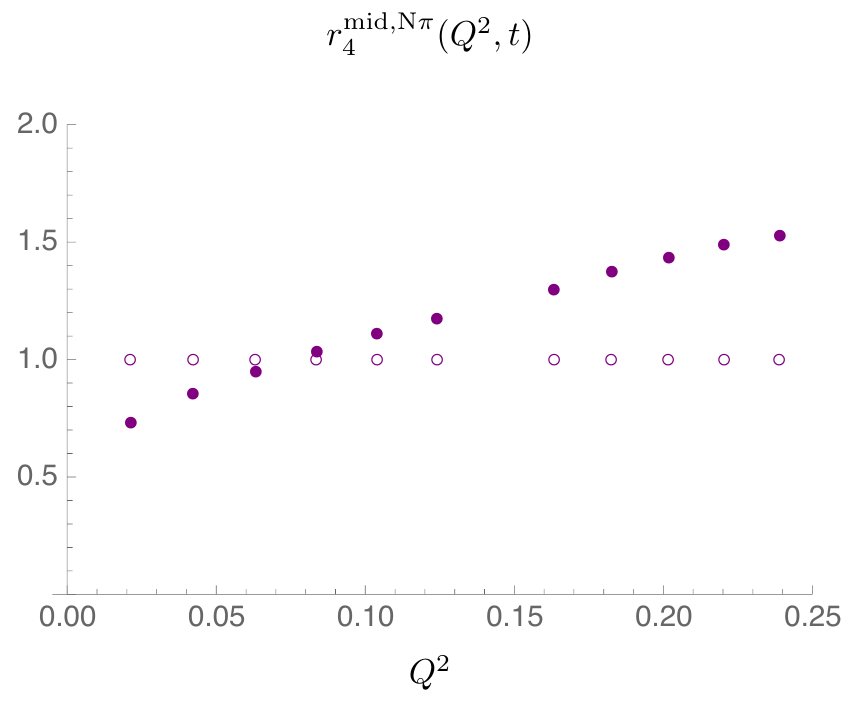}\\
\caption{\label{fig:r4}The ratio  $r^{\rm mid,\np}_{\rm 4}(Q^2,t)$, given in \pref{restNpi4}, for the lowest discrete momenta corresponding to $M_{\pi}L=6$ at $t=2$ fm (filled symbols) and for infinite $t$ (open symbols).}
\end{center}
\end{figure}

Figure \ref{fig:r4} shows the result for ratio $r^{{\rm est}, \np}_{\rm 4}(Q^2,t)$. The $\np$ contribution leads to a nearly linear dependence on $Q^2$ with non-vanishing positive slope.

%
\section{Comparison with recent PACS data}
%
 \subsection{Preliminaries}
 In order to compare the ChPT results for the form factor estimators with lattice QCD data we ideally need continuum extrapolated data with a (near to) physical pion mass. The spatial volume should be sufficiently large with $M_{\pi}L\gtrsim 4$, and the either the plateau or the midpoint estimates for the form factors should have been measured at sufficiently large euclidean time separations in the correlation functions. 

 In Ref.\ \cite{Ishikawa:2018rew} the PACS collaboration reports lattice data for the three nucleon form factors. The results were obtained in 2+1 flavor QCD on a $96^4$ lattice with lattice spacing $a\approx 0.085$ fm. Thus, the spatial lattice extent $L\approx 8.1$ fm is fairly large implying a small minimal pion momentum of about 155 MeV. The pion and nucleon masses are almost physical with $M_{\pi}\approx146$ MeV and $M_N\approx 958$ MeV. A fixed source-sink separation of 15 time slices has been used in the 3-pt functions, corresponding to $t\approx1.3$ fm, and the central four time slices were averaged to obtain the plateau estimates. For more simulation details see \cite{Ishikawa:2018rew}.

A source-sink separation of $t=1.3$ fm is very small, in fact too small to naively expect pion physics to dominate the correlation functions and ChPT to apply. Source-sink separations of 2 fm and larger are typically needed to sufficiently suppress the high-momentum $\np$ states that are not properly captured by ChPT.   
We nevertheless compare the ChPT results for the low-momentum $\np$ contribution with the PACS data, keeping in mind that the results are most probably subject to large corrections due to the neglected high momentum $\np$ and other excited states.  

 %
\subsection{The pseudoscalar form factor}
\begin{figure}[t]
\begin{center}
\includegraphics[scale=1.0]{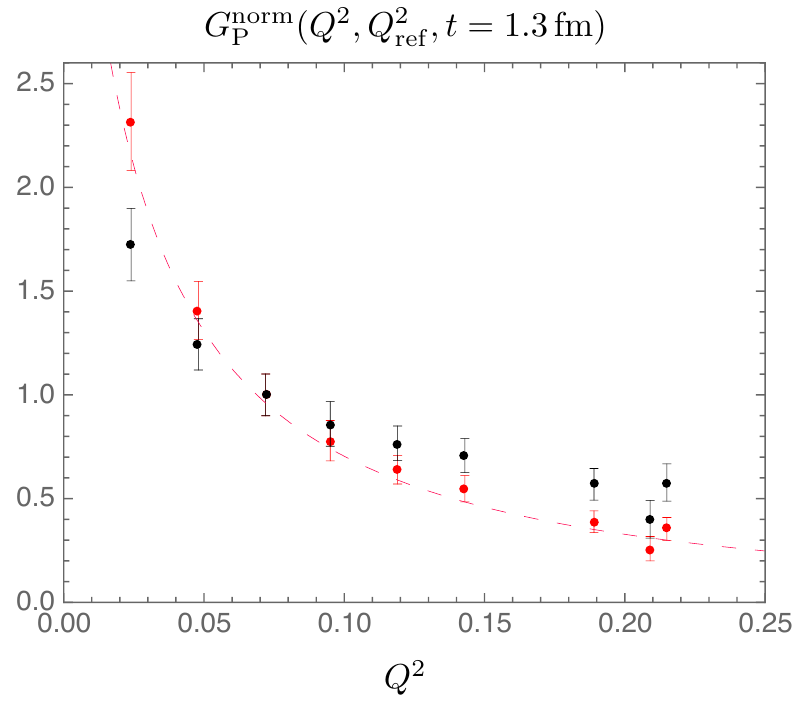}\\
\caption{\label{fig:GPnormPACS}PACS data for the normalized pseudoscalar form factor $\GP^{\rm norm}(Q^2,Q_{\rm ref}^2,t)$, defined in \pref{GPnorm},  for $t=1.3$ fm and $Q^2_{\rm ref} = 0.072\,({\rm GeV})^2$ (black symbols).  The dashed red line shows  the PPD result. Red symbols correspond to the data corrected with eq.\ \pref{pacsdatacorrected}. }
\end{center}
\end{figure}

%
Table IX of Ref.\ \cite{Ishikawa:2018rew} lists the plateau estimates of the pseudoscalar form factor for the nine lowest momentum transfers accessible in the simulation. The data are not renormalized and the renormalization factor $Z_{\rm P}$ is not yet available. $Z_m$ is known and could be used as an approximation for $Z_{\rm P}^{-1}$ \cite{Yamazakiprivcom, Ishikawa:2015fzw}, but we prefer to consider the normalized form factor 
\begin{equation}\label{GPnorm}
\GP^{\rm norm}(Q^2,Q_{\rm ref}^2,t)\equiv \frac{\GP^{\rm plat}(Q^2,t)}{\GP^{\rm plat}(Q_{\rm ref}^2,t)}\,,
\end{equation}
which is independent of $Z_{\rm P}$. Figure \ref{fig:GPnormPACS} shows this ratio (black symbols) for $Q^2_{\rm ref} = 0.072(2)\,({\rm GeV})^2$. The $Q^2$ dependence resembles the one of the induced pseudoscalar $\GPt(Q^2)$ with a strong $Q^2$ dependence at small momentum transfers. This is expected according to the PPD hypothesis. The red dashed line in figure \ref{fig:GPnormPACS} shows the ratio in \pref{GPnorm} with the PPD results \pref{ppd2} and \pref{GAdip} used on the right hand side.\footnote{We follow Ref.\ \cite{Ishikawa:2018rew} and set ${M}_{\rm A}\approx1.04$ GeV.}  Even though the statistical errors are quite large the momentum transfer dependence of the lattice data displays a flatter $Q^2$ dependence than the PPD model.

The plateau estimates were obtained at $t\approx 1.3$ fm, and we expect them to differ from the physical values at $t=\infty$ due to excited states. With the ChPT result $\epsilon^{\rm plat}_{\rm P}(Q^2,t)$  we can  analytically remove the anticipated LO $N\pi$-state contamination by calculating the corrected data
\begin{equation}\label{pacsdatacorrected}
\GP^{\rm corr}(Q^2,t) \equiv \frac{\GP^{\rm plat}(Q^2,t) }{1+ \epsilon^{\rm plat}_{\rm P}(Q^2,t)},
\end{equation}
setting $t=1.3$ fm. If higher order corrections and other excited-state contributions are small we expect
\begin{equation}\label{tindepofGPcorr}
\GP^{\rm corr}(Q^2,t) \approx \GP(Q^2)\,,
\end{equation}
i.e.\ the corrected data should be close to the true form factor. 
 
The red symbols in fig.\ \ref{fig:GPnormPACS} show the normalized form factor obtained with the corrected data on the right hand side of \pref{GPnorm}. Apparently, the corrected data show a steeper $Q^2$ dependence and are in better agreement with the PPD hypothesis. Taken at face value figure fig.\ \ref{fig:GPnormPACS} suggests that excited states other than low-momentum $\np$ states have only a small impact on the pseudoscalar form factor. This is surprising given the small source-sink separation.

\subsection{The ratios testing the gGT relation and the PPD hypothesis}

\begin{figure}[t]
\begin{center}
\includegraphics[scale=1.0]{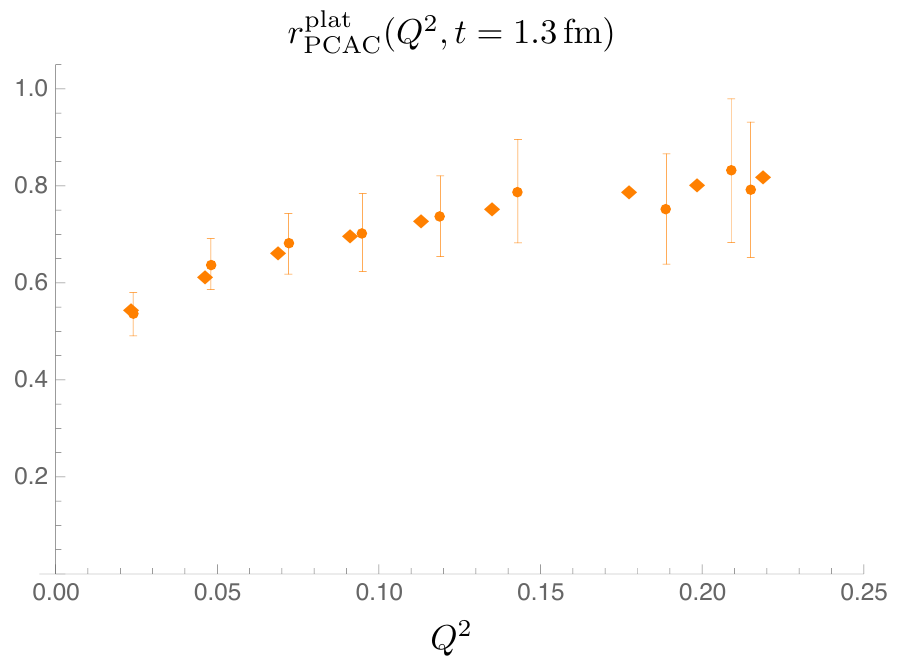}\\
\caption{\label{fig:ratiorPCAC} PACS data (circles) and the ChPT result (diamonds) for the ratio $r^{\rm plat}_{\rm PCAC}(Q^2,t=1.3 \, {\rm fm})$ given in \pref{restNpiPCAC} - \pref{restNpi2}.}
\end{center}
\end{figure}

Figure \ref{fig:ratiorPCAC} compares the PACS data (circles) and the ChPT result (diamonds) for the ratio $r^{\rm plat}_{\rm PCAC}$ that was introduced to test the gGT relation. We find good agreement within the (large) statistical errors, and the ChPT result describes very well the characteristic $Q^2$-dependent deviation of  $r^{\rm plat}_{\rm PCAC}$ from 1 observed in the lattice data. 

\begin{figure}[t]
\begin{center}
\includegraphics[scale=1.0]{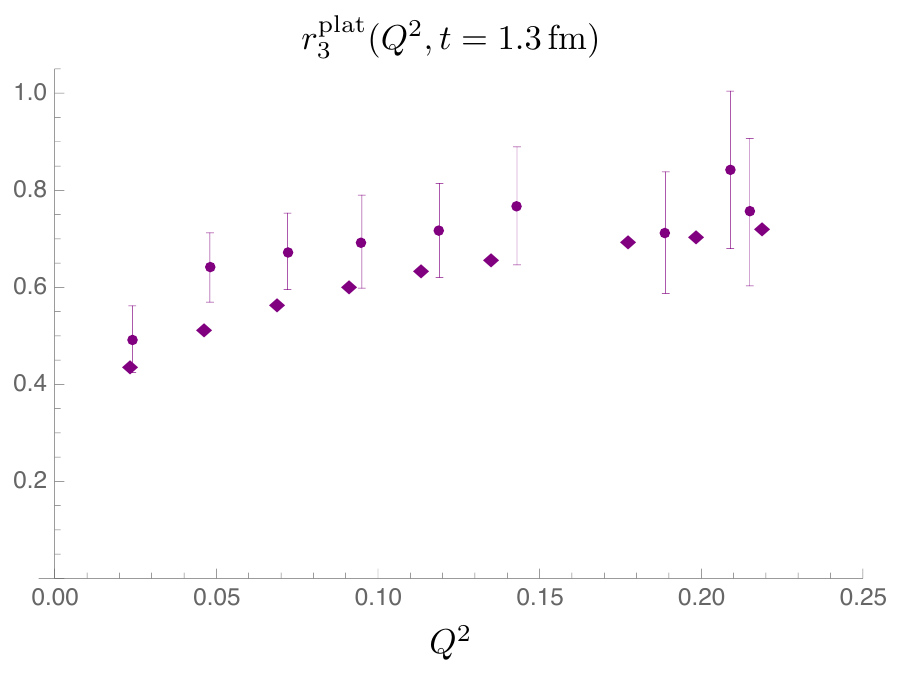}\\
\caption{\label{fig:ratioPPD} PACS data (circles) and the ChPT result (diamonds) for the ratio $r^{\rm plat}_{3}(Q^2,t=1.3 \, {\rm fm})$ given in \pref{restNpi3}.}
\end{center}
\end{figure}

\begin{figure}[t]
\begin{center}
\includegraphics[scale=1.0]{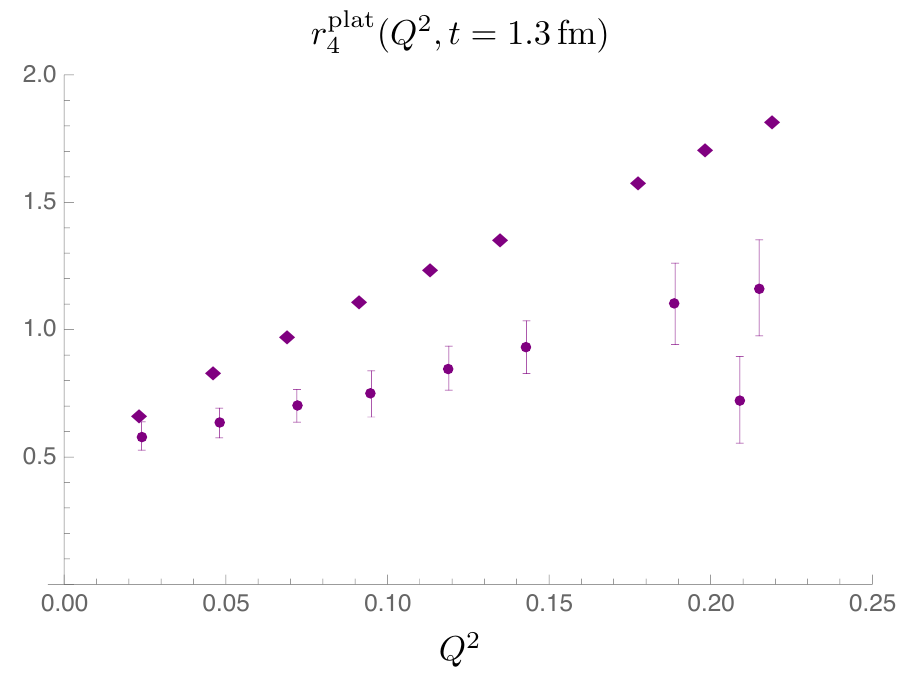}\\
\caption{\label{fig:ratiobarPPD} PACS data (circles) and the ChPT result (diamonds) for the ratio $r^{\rm plat}_{4}(Q^2,t=1.3 \, {\rm fm})$ given in \pref{restNpi4}.}
\end{center}
\end{figure}

Figures \ref{fig:ratioPPD} and \ref{fig:ratiobarPPD} show the PACS data  and the ChPT result for the two ratios ${r}^{\rm plat}_3$, ${r}^{\rm plat}_4$ that were introduced to test the PPD hypothesis. In case of ${r}^{\rm plat}_3$ we find good agreement, qualitatively similar to what is found for $r^{\rm plat}_{\rm PCAC}$. For the ratio ${r}^{\rm plat}_4$, on the the hand, the comparison is less satisfactory. Although we observe a monotonic increase as $Q^2$ gets larger, the slope in the lattice data is substantially smaller than in the ChPT result.

The discrepancy between the lattice data and the LO ChPT result displayed in figure \ref{fig:ratiobarPPD} is on the level one may expect for source-sink separations as small as 1.3 fm. As repeatedly said, at such a small time separation excited states other than low-momentum $\np$ states are not sufficiently suppressed to be negligible. Therefore, the difference between lattice data and ChPT may be attributed to excited state contributions not included in the ChPT result. More surprising and remarkable is the good agreement in figs.\ \ref{fig:ratiorPCAC} and \ref{fig:ratioPPD}, which seems to suggest that excited states other than low-momentum $\np$ states do not contribute significantly for the two ratios shown in these figures. 

For ${r}^{\rm plat}_{3}(Q^2,t)$ this is probably the case: The dominant source for the difference from 1 is the $\np$
contribution in the induced pseudoscalar form factor $\GPt$. As discussed in Ref.\ \cite{Bar:2018xyi}, the $\np$ contamination in $\GPt(Q^2,t)$ stems dominantly from a single state with the spatial pion momentum fixed to the one associated with the momentum transfer $Q^2$. The contribution of other $\np$ states with increasing nucleon and pion momenta essentially cancel  in the sum over all $\vec{k}$. Such a (partial) cancellation does not happen in case of the $\np$ contribution to $\GP$. Therefore, ChPT is expected to work less well for ${r}^{\rm plat}_{4}$. Finally, the $\np$ contamination in $\GPt$ dominates the ratio $r^{\rm plat}_{\rm PCAC}$, in particular for large $Q^2$. Thus, the good agreement we observe in fig.\ \ref{fig:ratiorPCAC} is basically the same agreement we see in fig.\ \ref{fig:ratioPPD}. 

\section{Discussion}

It is premature to draw definite conclusions from the comparison in the last section. More data is needed to unambiguously identify $\np$ excited states as the dominant source for the observed violation of the gGT relation. To this end lattice data for larger and more than one source-sink separation will be very beneficial, since these allow for checking the dependence of $\epsilon_{\rm X}(Q,t)$ on the source-sink separation $t$.  The PACS collaboration reported on form factor data  obtained from an ensemble with a 135 MeV pion mass and a finite volume of size $(10.8 {\rm \, fm})^4$  \cite{Shintani:2018ozy}. Plateau estimates for the form factors exist for four source-sink separations, and a dependence on the source-sink separation is clearly visible in the data.  It is highly interesting to compare the data with the ChPT predictions presented here, but the data are not publicly available yet \cite{EigoShintaniPC}.

Although suggestive, attributing the violation of the gGT relation to $\np$ states leads to an apparent puzzle. Many lattice collaborations employ multi-state fits to control the anticipated excited-state contamination.  Ref.\ \cite{Rajan:2017lxk} reports on fits to the 3-pt functions data including two excited states next to the ground state. Still, $r_{\rm PCAC}$ is significantly smaller than 1 and looks qualitatively the same compared to the ratio formed with simple plateau estimates for the form factors. If the violation of the gGT relation is due to the excited-state contamination one may legitimately ask why multi-state fits are unable to capture it properly \cite{RajanGuptaPC}.

The ChPT results for the $\np$-state contamination offer a possible answer to this question. It is necessary to recall how most multi-state fits are done in practice. To extract the form factors from the 3-pt correlation function one considers the spectral decomposition including the ground state and one or two excited states. The energies of these states are usually taken from fits to the nucleon 2-pt function, since the contributing states are expected to be the same. 

Although correct in principle, this assumption might be flawed in some cases.
The spectral decomposition for the 2-pt function is given in \pref{2ptDecomp2}. For physical pion masses and on typical volumes with $M_{\pi}L\gtrsim 4$  quite a few  $\np$-states contribute non-negligibly to the sum in \pref{2ptDecomp2}, not only one \cite{Bar:2017kxh}.
Since lattice data have statistical errors one may still be able to fit a 2-state fit ansatz,
\begin{eqnarray}\label{2ptDecompeff}
C_2(\vec{q},t) & \approx & C^N_2(\vec{q},t)\left\{1+  d^{\rm eff}(\vec{q})e^{-\Delta E^{\rm eff}(\vec{q}) t}\right\}\label{DefC2eff}\,,
\end{eqnarray}
to the 2-pt function data with effective parameters $d^{\rm eff}(\vec{q})$ and $\Delta E^{\rm eff}(\vec{q})$. Both
will be some average of the contributing coefficients $d(\vec{q},\vec{k})$ and energy gaps $\Delta E(\vec{q},\vec{k})$, respectively. Since the coefficients $d(\vec{q},\vec{k})$ are all positive numbers the average energy gap $\Delta E^{\rm eff}(\vec{q})$ will be larger than the lowest or even a few individual gaps $\Delta E(\vec{q},\vec{k})$. For simplicity we have assumed one excited state in \pref{2ptDecompeff}, but the same arguments apply for more than one effective energy gap if an $n$-state ansatz is made with $n>1$.

The analogous spectral decomposition of the 3-pt function is given in \pref{C3muspecDecomp} - \pref{DefZmu}, and the same energy gaps appear in both cases. However, suppose we are interested in calculating the induced pseudoscalar form factor $\GPt$. It is directly proportional to the ratio $R_{\mu}(\vec{q},t,t')$ with $\mu=1$, as long as the momentum transfer $\vec{q}$ can be chosen with both components $q_1$ and $q_3$ non-vanishing, see eq.\ \pref{AsympValueR33}. For this particular case the coefficients $b_{1}(\vec{q},\vec{k}), \tilde{b}_{1}(\vec{q},\vec{k})$ and $c_{1}(\vec{q},\vec{k})$ are proportional to the product $k_1k_3/q_1q_3$ \cite{Bar:2018xyi}.
Therefore, performing the sum over all pion momenta $\vec{k}$ their contributions in \pref{DefZmu} essentially cancel out. Consequently, to a good approximation the $\np$-state contribution in the 3-pt function reads \cite{Bar:2018xyi}
\begin{equation}\label{approx}
Z_{1}(\vec{q},t,t')  \approx  a_{1}(\vec{q}) e^{-\Delta E(0,\vec{q}) (t-t')}+ \tilde{a}_{1}(\vec{q}) e^{-\Delta E(\vec{q},-\vec{q})t'}\,.
\end{equation}
Both gaps in here are approximately equal to $E_{\pi,\vec{q}}$, the energy of a pion with spatial momentum associated with the momentum transfer. The larger the spatial volume the smaller can $|\vec{q}|$ be, and the larger one can expect the difference between the effective gap from the 2-pt function and the gap in the 3-pt function.\footnote{In the 3-pt function with $\mu=3$ the contributions proportional to $b_{3}(\vec{q},\vec{k}), \tilde{b}_{3}(\vec{q},\vec{k})$ and $c_{3}(\vec{q},\vec{k})$ are proportional to $k_3^2$, and do not cancel when the sum is performed \cite{Bar:2018xyi}.} 

How exactly the lattice estimate for $\GPt$ is influenced by a misidentified energy gap in the 3-pt function is an open question. And even if it has an impact it is not clear whether it provides the answer to the question why multi-state fits are apparently unable to capture the excited-state contribution in some cases. Nevertheless, since the $\np$-state contamination in the induced pseudoscalar form factor dominates the ratio $r_{\rm PCAC}$ it is conceivable that it is at least part of the puzzle. 

The same conclusion has been put forward in \cite{Jang:2019vkm}. It is shown in this reference that the energy gap to the first excited state in the 3-pt function with the temporal component $A_4$ of the axial vector is different from the one in the nucleon 2-pt function. This too is predicted by the ChPT results in Ref.\ \cite{Bar:2018xyi}, the argument being essentially the same as we have given above for \pref{approx}. 
It will be very interesting to repeat the study in \cite{Jang:2019vkm} for the component  $A_1$ of the axial vector. This hopefully will shed additional light on the source for the violation of the gGT relation.

Recently, Ref.\ \cite{Bali:2018qus} suggested the use of a projected pseudoscalar density $P^{\perp}(x)$ as a method to reduce the excite state contamination. The efficiency of this projection in removing the $N\pi$-state contamination can also be studied analytically in ChPT. First results to leading order in the NR expansion suggest that the $N\pi$ contamination in the correlation functions with $P^{\perp}(x)$ are even larger compared to the case with the standard pseudoscalar density $P(x)$ discussed here. This will be discussed in a forthcoming publication  \cite{OBWiP}.

\vspace{4ex}
\noindent {\bf Acknowledgments}
\vspace{2ex}

Discussions and/or correspondence with C.\ Alexandrou, R.~Gupta,  Y.-C.~Jang, R.~Sommer,  and T.~Yamazaki is gratefully acknowledged.
This work was supported by the German Research Foundation (DFG), Grant ID BA 3494/2-1.

\end{document}